\documentclass
[superscriptaddress,nobibnotes,aps,prd,nofootinbib]{revtex4}%
\usepackage{amsmath}
\usepackage{amsfonts}
\usepackage{amssymb}
\usepackage{graphicx}%
\begin{document}
\title{Generalized Absurdly Benign Traversable Wormholes powered by Casimir Energy}
\author{Remo Garattini}
\email{Remo.Garattini@unibg.it}
\affiliation{Universit\`{a} degli Studi di Bergamo,\\
Dipartimento di Ingegneria e Scienze Applicate,\\
Viale Marconi 5, 24044 Dalmine (Bergamo) Italy and\\
I.N.F.N. - sezione di Milano, Milan, Italy.}

\begin{abstract}
In this work, we explore the connection between Casimir energy and an Absurdly
Benign Traversable Wormhole, which in the literature has been considered only
in the pioneering paper of Morris and Thorne. To have consistency with the
Casimir source, we need to generalize the idea of an Absurdly Benign
Traversable Wormhole into a Generalized Absurdly Benign Traversable Wormhole.
With this generalization, we have found that the wormhole throat is not more
Planckian, but huge. Three profiles have been studied: one of them is directly
connected with the Casimir source, while the other two have been obtained
approximating the first one close to the throat. In all profiles the wormhole
throat size is predicted to be of the order of $10^{17}m$. This huge size can
be fine tuned by modulating the original Casimir energy source size. We have
also found that the traceless and divergenceless property of the original
Casimir stress energy tensor is here partially reproduced.

\end{abstract}
\maketitle

\section{Introduction}

GW150914 is the acronym associated to the first ever detection of
gravitational waves from the merger of two black holes. Advanced LIGO made
this first observation during the period running from 12th September 2015 to
19th January 2016 \cite{GW150914} and its second run from 30th November 2016
and 25th August 2017. What is the meaning of such an observation? The meaning
is that another prediction of General Relativity has been confirmed and this
shows that General Relativity is a quite solid theory for gravitation. Among
other things, such a theory predicts other interesting objects: Traversable
Wormholes. Traversable Wormholes (TW)\ are solutions of the Einstein's Field
Equations and even if there is no experimental evidence of their existence,
their interest has been growing because potentially they can be considered as
black hole mimickers \cite{CFP}. The interesting feature of a traversable
wormhole is its ability in connecting remote regions of space-time in a
reasonable amount of time: of course reasonable compared with the duration of
human life \cite{MT,MTY,Visser}. Unfortunately, the traversability is
accompanied by unavoidable violations of null energy conditions, namely, the
matter threading the wormhole's throat has to be exotic. Classical matter
satisfies the usual energy conditions. Therefore, it is likely that wormholes
must belong to the realm of semiclassical or perhaps a possible quantum theory
of the gravitational field. In the context of Quantum Field Theory, a possible
source of exotic matter could be represented by the Casimir energy
\cite{Casimir}. The Casimir energy appears between two plane parallel, closely
spaced, uncharged, metallic plates in vacuum. This phenomenon develops an
attractive force; it was predicted theoretically in 1948 and experimentally
confirmed in the Philips laboratories \cite{Sparnaay}. However, only in recent
years further reliable experimental investigations have confirmed such a
phenomenon \cite{Lamoreaux,BCOR}. As far as we know, the Casimir energy
represents the only artificial source of negative energy, whose value in terms
of energy density is%
\begin{equation}
\rho_{0}=-\frac{\hbar c\pi^{2}}{720d^{4}}.\label{rhoC}%
\end{equation}
Its Stress-Energy Tensor (SET) is represented by%
\begin{equation}
T_{\mu\nu}=\frac{\hbar c\pi^{2}}{720d^{4}}\left[  diag\left(
-1,-3,1,1\right)  \right]  ,\label{Tmn}%
\end{equation}
where $d$ is the separation of the plates, and has the following properties:
it is divergenceless and traceless. Nevertheless, where is the connection
between a TW and the Casimir energy. First of all to describe a spherically
symmetric and static TW, we need to introduce a spacetime metric of the form%
\begin{equation}
ds^{2}=-e^{2\Phi(r)}\,dt^{2}+\frac{dr^{2}}{1-b(r)/r}+r^{2}\,(d\theta^{2}%
+\sin^{2}{\theta}\,d\varphi^{2})\,,\label{ds}%
\end{equation}
where $\Phi(r)$ and $b(r)$ are arbitrary functions of the radial coordinate,
$r$, denoted as the redshift function, and the shape function, respectively
\cite{MT,Visser}. The radial coordinate has a range that increases from a
minimum value at $r_{0}$, corresponding to the wormhole throat, to infinity. A
fundamental property of a wormhole is that a flaring out condition of the
throat, given by $(b-b^{\prime}r)/b^{2}>0$, is imposed \cite{MT,Visser}, and
at the throat $b(r_{0})=r_{0}$, the condition $b^{\prime}(r_{0})<1$ is imposed
to have wormhole solutions. Another condition that needs to be satisfied is
$1-b(r)/r>0$. For the wormhole to be traversable, one must demand that there
are no horizons present, which are identified as the surfaces with $e^{2\phi
}\rightarrow0$, so that $\phi(r)$ must be finite everywhere. Using the
Einstein Field Equations (EFE) $\left(  \kappa=8\pi G/c^{4}\right)  $%
\begin{equation}
G_{\mu\nu}=\kappa T_{\mu\nu},
\end{equation}
in an orthonormal reference frame, we obtain the following set of equations%
\begin{equation}
\rho\left(  r\right)  =\frac{1}{\kappa}\frac{b^{\prime}}{r^{2}},\label{rho}%
\end{equation}%
\begin{equation}
p_{r}\left(  r\right)  =\frac{1}{\kappa}\left[  \frac{2}{r}\left(
1-\frac{b\left(  r\right)  }{r}\right)  \Phi^{\prime}-\frac{b}{r^{3}}\right]
,\label{pr}%
\end{equation}%
\begin{align}
p_{t}(r) &  =\frac{1}{\kappa}\Bigg\{\left(  1-\frac{b}{r}\right)  \left[
\Phi^{\prime\prime}+\Phi^{\prime}\left(  \Phi^{\prime}+\frac{1}{r}\right)
\right]  \nonumber\\
&  -\frac{b^{\prime}r-b}{2r^{2}}\left(  \Phi^{\prime}+\frac{1}{r}\right)
\Bigg\},
\end{align}
in which $\rho\left(  r\right)  $ is the energy density, $p_{r}\left(
r\right)  $ is the radial pressure, and $p_{t}\left(  r\right)  $ is the
lateral pressure. Using the conservation of the stress-energy tensor, in the
same orthonormal reference frame, we get%
\begin{equation}
p_{r}^{\prime}=\frac{2}{r}\left(  p_{t}-p_{r}\right)  -\left(  \rho
+p_{r}\right)  \Phi^{\prime}.
\end{equation}
Finally, the EFE can be rearranged to give%
\begin{equation}
b^{\prime}=\kappa\rho\left(  r\right)  r^{2},\label{EFE1}%
\end{equation}%
\begin{equation}
\Phi^{\prime}=\frac{b+\kappa p_{r}r^{3}}{2r^{2}\left(  1-\frac{b\left(
r\right)  }{r}\right)  }.
\end{equation}
However, given the quantum nature of the Casimir effect, the EFE must be
replaced with the semiclassical EFE, namely%
\begin{equation}
G_{\mu\nu}=\kappa\left\langle T_{\mu\nu}\right\rangle ^{\text{Ren}},
\end{equation}
where $\left\langle T_{\mu\nu}\right\rangle ^{\text{Ren}}$ describes the
renormalized quantum contribution of some matter fields: in this specific
case, the electromagnetic field. In a recent paper \cite{EPJC}, a connection
between the SET $\left(  \ref{Tmn}\right)  $ and the spacetime metric $\left(
\ref{ds}\right)  $ has been deduced, obtaining a line element of the form%
\begin{equation}
ds^{2}=-{\left(  \frac{3r}{3r+r_{0}}\right)  }^{2}\,dt^{2}+\frac{dr^{2}%
}{1-\frac{2r_{0}}{3r}-\frac{r_{0}^{2}}{3r^{2}}}+r^{2}\,(d\theta^{2}+\sin
^{2}{\theta}\,d\varphi^{2})\,,\label{dSC3}%
\end{equation}
with%
\begin{equation}
\phi\left(  r\right)  =\ln\left(  \frac{3r}{3r+r_{0}}\right)  \qquad
and\qquad{b}\left(  r\right)  =\frac{2r_{0}}{3}+\frac{r_{0}^{2}}%
{3r}.\label{phi(r)b(r)}%
\end{equation}
Such a result has been obtained promoting the plate separation $d$ in the SET
to a variable $r$ and an Equation of State (EoS) of the form $p_{r}\left(
r\right)  =\omega\rho\left(  r\right)  $ has been used with $\omega=3$. A
further investigation about the line element $\left(  \ref{dSC3}\right)  $ has
been done in Refs.\cite{Weak,GUP,SKT}. The purpose of this paper is to
establish if other connections between the Casimir energy and a TW can be
created. The motivation to insist in this research is dictated by the fact
that a TW needs exotic matter and at the moment only the Casimir energy can be
used as an appropriate source. However, this time the plate separation $d$
will be considered as fixed and not a variable. Surprisingly, we will obtain
an interesting result even in this configuration. The rest of the paper is
structured as follows, in section \ref{p1} we continue the investigation to
determine if the Casimir energy density $\left(  \ref{rhoC}\right)  $ can be
considered as a source for an Absurdly Benign Traversable Wormhole, in section
\ref{p2} we generalize the definition of an Absurdly Benign Traversable
Wormhole, in section \ref{p3} we give another profile for the Generalized
Absurdly Benign Traversable Wormhole directly connected with the Casimir
source $\left(  \ref{rhoC}\right)  $ but without approximation. We summarize
and conclude in section \ref{p4}. Units in which $\hbar=c=k=1$ are used
throughout the paper and will be reintroduced when they are relevant.

\section{The Absurdly Benign Traversable Wormhole and the Equation of State}

\label{p1}The original case, that in which, the plate separation $d$ is fixed,
has an energy density expressed by Eq. $\left(  \ref{rho}\right)  $. With the
help of Eq.$\left(  \ref{EFE1}\right)  $, we can compute the form of the shape
function%
\begin{equation}
b\left(  r\right)  =r_{0}-\frac{\pi^{3}}{270d^{4}}\left(  \frac{\hbar G}%
{c^{3}}\right)  \left(  r^{3}-r_{0}^{3}\right)  , \label{b(r)C}%
\end{equation}
which is not asymptotically flat but asymptotically de Sitter. Indeed, the
Casimir energy acts like a \textquotedblleft\textit{Cosmological
Constant}\textquotedblright. This means that a TW in a strict sense cannot be
formed. On this point we will come back in Section \ref{p4}. However, in
proximity of the throat, the shape function $\left(  \ref{b(r)C}\right)  $ can
be rearranged in the following way%
\begin{align}
b\left(  r\right)   &  =r_{0}-\frac{\pi^{3}}{270d^{4}}\left(  \frac{\hbar
G}{c^{3}}\right)  \left(  r-r_{0}\right)  \left(  r^{2}+r_{0}r+r_{0}%
^{2}\right) \nonumber\\
&  \simeq r_{0}\left(  1-\frac{r_{0}\pi^{3}}{90d^{4}}\left(  \frac{\hbar
G}{c^{3}}\right)  \left(  r-r_{0}\right)  \right)  =r_{0}\left(  1-\frac
{r_{0}l_{P}^{2}\pi^{3}}{90d^{4}}\left(  r-r_{0}\right)  \right)  .
\label{b(r)Cr0}%
\end{align}
The shape function $\left(  \ref{b(r)Cr0}\right)  $ has the same structure of
an \textit{Absurdly Benign Traversable Wormhole }(ABTW) proposed by Morris and
Thorne in Ref.\cite{MT} except for the exponent. I recall to the reader that
an ABTW is defined in such a way to have exotic matter concentrated into the
region $r_{0}\leq r\leq r_{0}+a$ with $a/r_{0}\ll1$. In practice, the shape
function and the redshift functions are defined by%
\begin{align}
b(r)  &  =r_{0}\left(  1-\left(  \frac{r-r_{0}}{a}\right)  \right)
^{2},\qquad\Phi(r)=0;\qquad r_{0}\leq r\leq r_{0}+a\nonumber\\
b(r)  &  =0,\qquad\Phi(r)=0;\qquad r\geq r_{0}+a. \label{b(r)AB}%
\end{align}
Therefore outside the location $r=r_{0}+a$, the spacetime is Minkowski. If we
make the identification%
\begin{equation}
a=\frac{90d^{4}}{r_{0}l_{P}^{2}\pi^{3}}, \label{a}%
\end{equation}
then the similarity between $\left(  \ref{b(r)C}\right)  $ and $\left(
\ref{b(r)AB}\right)  $ improves and if we put real numbers coming from
experiments, we find%
\begin{equation}
a\simeq\frac{90\left(  10^{-9}m\right)  ^{4}}{r_{0}\pi^{3}\left(
1.6\times10^{-35}m\right)  ^{2}}\simeq\frac{10^{34}m^{2}}{r_{0}}, \label{aa}%
\end{equation}
where we have assumed a plate separation of the order of the $nm$, which is
the actual distance used in laboratories. In order to have $a$ small, we have
two possibilities:

\begin{description}
\item[a)] either $r_{0}\gg$ $10^{34}m$

\item or

\item[b)] the plate separation $d$ must be less of the order of a $fm$ to have
a wormhole throat of the order of $10^{10}m$, which is bigger of the diameter
of the sun.
\end{description}

In both cases, the use of such a source is neither practical nor physically
meaningful. If condition a) or b) are not satisfied, the exotic matter is not
concentrated close to the throat, rather is distributed in a wide area of the
space. Therefore, it appears important to establish if a better connection
between the Casimir energy and an ABTW exists in order to use such a source.
Note that except for Refs.\cite{MT,Visser} and Ref.\cite{JFW}, the subject of
ABTW has not examined extensively. To further proceed, we introduce an
inhomogeneous Equation of State (EoS) of the form%
\begin{equation}
p_{r}\left(  r\right)  =\omega\left(  r\right)  \rho\left(  r\right)  .
\label{Inhom}%
\end{equation}
From Eqs.$\left(  \ref{rho}\right)  $ and $\left(  \ref{pr}\right)  $, by
imposing%
\begin{equation}
b\left(  r\right)  +\kappa p_{r}\left(  r\right)  r^{3}=0, \label{constr}%
\end{equation}
we find%
\begin{equation}
\omega\left(  r\right)  =-\frac{b\left(  r\right)  }{b^{\prime}\left(
r\right)  r} \label{o(r)}%
\end{equation}
and the EFE can be solved to give%
\begin{equation}
b(r)=r_{0}\,\exp\left[  -\int_{r_{0}}^{r}\,\frac{d\bar{r}}{\omega(\bar{r}%
)\bar{r}}\right]  \,. \label{form}%
\end{equation}
From the shape function $\left(  \ref{form}\right)  $, we can compute
$\rho\left(  r\right)  $%
\begin{equation}
\rho\left(  r\right)  =-\frac{r_{0}}{\kappa r^{3}\omega\left(  r\right)
}\,\exp\left[  -\int_{r_{0}}^{r}\,\frac{d\bar{r}}{\omega(\bar{r})\bar{r}%
}\right]  , \label{rho1}%
\end{equation}
$p_{r}\left(  r\right)  $%
\begin{equation}
p_{r}\left(  r\right)  =-\frac{r_{0}}{\kappa r^{3}}\exp\left[  -\int_{r_{0}%
}^{r}\,\frac{d\bar{r}}{\omega(\bar{r})\bar{r}}\right]  \label{pr1}%
\end{equation}
and $p_{t}\left(  r\right)  $%
\begin{equation}
p_{t}(r)=\frac{r_{0}}{2\kappa r^{3}}\left(  \frac{1}{\omega\left(  r\right)
}+1\right)  \exp\left[  -\int_{r_{0}}^{r}\,\frac{d\bar{r}}{\omega(\bar{r}%
)\bar{r}}\right]  . \label{pt1}%
\end{equation}
If the relationship $\left(  \ref{o(r)}\right)  $ is satisfied, then we have a
zero-tidal-force wormhole (ZTF), a condition represented by the choice
$\Phi(r)=0$. This is also the same condition assumed for the ABTW. For
instance, for the ABTW, it is immediate to obtain that%
\begin{gather}
\omega\left(  r\right)  =\frac{a}{2r}\left(  1-\left(  \frac{r-r_{0}}%
{a}\right)  \right)  ,\nonumber\\
\omega\left(  r_{0}\right)  =\frac{a}{2r_{0}}\qquad\qquad\omega\left(
r_{0}+a\right)  =0.
\end{gather}
If $0>\omega\left(  r\right)  >-1$, we are in the Dark Energy sector, while if
$-1>\omega\left(  r\right)  $, then we are in the Phantom Energy
sector\footnote{Note that the energy density of the SET $\left(
\ref{Tmn}\right)  $ has been considered in the context of phantom energy in
Ref.\cite{Sushskov}.}. To complete the discussion we include also the form of
the SET%
\begin{gather}
T_{\mu\nu}=\frac{r_{0}}{\kappa r^{3}}diag\left(  -\frac{1}{\omega\left(
r\right)  },-1,\frac{1}{2\omega\left(  r\right)  }+\frac{1}{2},\frac
{1}{2\omega\left(  r\right)  }+\frac{1}{2}\right)  \exp\left[  -\int_{r_{0}%
}^{r}\,\frac{d\bar{r}}{\omega(\bar{r})\bar{r}}\right] \nonumber\\
=-\frac{b(r)}{\kappa r^{3}\omega\left(  r\right)  }diag\left(  1,\omega\left(
r\right)  ,-\frac{1}{2}-\frac{\omega\left(  r\right)  }{2},-\frac{1}{2}%
-\frac{\omega\left(  r\right)  }{2}\right) \nonumber\\
=\rho\left(  r\right)  diag\left(  1,\omega\left(  r\right)  ,-\frac{1}%
{2}-\frac{\omega\left(  r\right)  }{2},-\frac{1}{2}-\frac{\omega\left(
r\right)  }{2}\right)  . \label{SET}%
\end{gather}
By construction the SET $\left(  \ref{SET}\right)  $ is divergenceless, but it
is not traceless. However, it is always possible to rearrange the previous SET
$\left(  \ref{SET}\right)  $ in such a way to extract the traceless part.
Indeed%
\begin{equation}
T_{\mu\nu}=T_{\mu\nu}^{T}+\frac{T}{4}g_{\mu\nu}=\frac{\rho\left(  r\right)
}{2}\left[  diag\left(  1,2\omega\left(  r\right)  +1,-\omega\left(  r\right)
,-\omega\left(  r\right)  \right)  -g_{\mu\nu}\right]  , \label{TmnDec}%
\end{equation}
where $T_{\mu\nu}^{T}$ is the traceless part of the SET $\left(
\ref{SET}\right)  $. It is interesting to observe that by imposing the
following condition%
\begin{equation}
\omega\left(  r_{0}\right)  =1, \label{constr1}%
\end{equation}
one finds that%
\begin{equation}
T_{\mu\nu}^{T}=\frac{\rho\left(  r_{0}\right)  }{2}\left[  diag\left(
1,3,-1,-1\right)  \right]  , \label{TTmn}%
\end{equation}
independently on the form of $\omega\left(  r\right)  $. In the expression
$\left(  \ref{TTmn}\right)  $, we can recognize the Casimir structure of the
SET. In the next section, we will examine some specific profiles of
$\omega\left(  r\right)  $ which can lead to a generalized ABTW (GABTW).

\section{The Generalized Absurdly Benign Traversable Wormhole}

\label{p2}In section \ref{p1}, we have derived a form for the shape function
directly by the Casimir energy density described by Eq.$\left(  \ref{b(r)C}%
\right)  $. We have also given the expression in proximity of the throat
described by Eq.$\left(  \ref{b(r)Cr0}\right)  $ and we have deduced that it
seems to have a connection with the ABTW. However, the connection is not
complete, because from Eq.$\left(  \ref{b(r)Cr0}\right)  $, we can easily
derive the energy density which has the following form%
\begin{equation}
\rho\left(  r\right)  =-\frac{r_{0}^{2}l_{P}^{2}\pi^{3}}{90d^{4}\kappa r^{2}%
}=-\frac{r_{0}^{2}\pi^{2}\hbar c}{720d^{4}r^{2}}\underset{r=r_{0}%
}{\Longrightarrow}-\frac{\pi^{2}\hbar c}{720d^{4}}=-\rho_{0},
\end{equation}
which is exactly the Casimir energy density only on the throat, as it should
be. Nevertheless, if we assume the validity of the identification $\left(
\ref{a}\right)  $, then the associated SET is not the Minkowski SET outside
the point $\bar{r}=r_{0}+a$, but it has the following expression%
\begin{equation}
T_{\mu\nu}=\rho_{0}\frac{r_{0}^{2}}{\bar{r}^{2}}diag\left(  -1,0,\frac{1}%
{2},\frac{1}{2}\right)  . \label{TmnO}%
\end{equation}
Furthermore, the assumption $\left(  \ref{a}\right)  $ is physically
meaningless. Therefore, we are going to explore different profiles with the
hope they satisfy as much as possible the ABTW form and, at the same time, the
form of the Casimir SET. Before going on we have to observe one point: the
ABTW shape function is%
\begin{equation}
b(r)=r_{0}\left(  1-\left(  \frac{r-r_{0}}{a}\right)  \right)  ^{2}%
\end{equation}
and close to the throat, the leading term is%
\begin{equation}
b(r)\simeq r_{0}\left(  1-\frac{2}{a}\left(  r-r_{0}\right)  \right)  ,
\end{equation}
which has the same form of \ref{b(r)Cr0}. This means that the ABTW can be the
right prototype to build a better profile. This tentative improvement will be
done by means of the inhomogeneous EoS $\left(  \ref{Inhom}\right)  $ which,
in the case of the ABTW obeys the EoS $\left(  \ref{Inhom}\right)  $ with the
following relationship%
\begin{equation}
\omega\left(  r\right)  =\frac{1-\mu\left(  r-r_{0}\right)  }{2\mu r}%
;\qquad\mu=\frac{1}{a}. \label{o(r)AB}%
\end{equation}
As a first proposal, we will examine the following profile

\subsection{Example I $\omega\left(  r\right)  =\left(  1-\mu\left(
r-r_{0}\right)  \right)  /\alpha\mu r$}

When $\omega\left(  r\right)  $ assumes the following profile%
\begin{equation}
\omega\left(  r\right)  =\frac{1-\mu\left(  r-r_{0}\right)  }{\alpha\mu
r};\qquad\alpha>1, \label{o(r)ABG}%
\end{equation}
which is an immediate generalization of the relationship $\left(
\ref{o(r)AB}\right)  $, one finds%
\begin{equation}
\omega\left(  r\right)  \rightarrow\left\{
\begin{array}
[c]{cc}%
1/\alpha\mu r_{0} & r\rightarrow r_{0}\\
-1/\alpha & r\rightarrow\infty
\end{array}
\right.  ;\qquad r\in\left[  r_{0},\infty\right)  ,
\end{equation}
where $\mu$ is an inverse length scale. However, it is of much more interest
the following assumption%
\begin{equation}
\omega\left(  r\right)  =0\qquad\mathrm{when}\qquad r=\bar{r}=r_{0}+\frac
{1}{\mu}. \label{o(r)GABTW0}%
\end{equation}
Note that outside the point $r=\bar{r}$, if we not impose $\omega\left(
r\right)  =0$, we are in the dark energy sector because $0>\omega\left(
r\right)  >-1$, since $\alpha>1$. When choice $\left(  \ref{o(r)GABTW0}%
\right)  $ is adopted, from Eq.$\left(  \ref{form}\right)  $, one finds%
\begin{align}
b(r)  &  =r_{0}\,\exp\left[  -\int_{r_{0}}^{r}\,\frac{\alpha\mu\bar{r}d\bar
{r}}{1-\mu\left(  \bar{r}-r_{0}\right)  \bar{r}}\right]  =r_{0}\,\exp\left[
-\alpha\mu\int_{r_{0}}^{r}\,\frac{d\bar{r}}{1-\mu\left(  \bar{r}-r_{0}\right)
}\right] \nonumber\\
&  =r_{0}\,\exp\left[  \alpha\ln\left(  1-\mu\left(  r-r_{0}\right)  \right)
\right]  =r_{0}\left(  1-\mu\left(  r-r_{0}\right)  \right)  ^{\alpha}.
\end{align}
As we can see, the choice $\left(  \ref{o(r)ABG}\right)  $ leads to a
generalized ABTW, if we adopt also the following further conditions%
\begin{align}
b(r)  &  =r_{0}\left(  1-\mu\left(  r-r_{0}\right)  \right)  ^{\alpha}%
,\qquad\Phi(r)=0;\qquad r_{0}\leq r\leq r_{0}+1/\mu\nonumber\\
b(r)  &  =0,\qquad\Phi(r)=0;\qquad r\geq r_{0}+1/\mu. \label{b(r)ABG}%
\end{align}
Then the components of the stress-energy tensor can be easily computed to
obtain%
\begin{align}
\rho\left(  r\right)   &  =-\frac{r_{0}\alpha\mu}{\kappa r^{2}}\left(
1-\mu\left(  r-r_{0}\right)  \right)  ^{\alpha-1},\label{rhoABG}\\
p_{r}\left(  r\right)   &  =-\frac{r_{0}\left(  1-\mu\left(  r-r_{0}\right)
\right)  ^{\alpha}}{\kappa r^{3}},\label{prABG}\\
p_{t}(r)  &  =\frac{r_{0}\left(  1-\mu\left(  r-r_{0}\right)  \right)
^{\alpha-1}}{2\kappa r^{3}}\left(  1-\mu\left(  r\left(  1-\alpha\right)
-r_{0}\right)  \right)  , \label{ptABG}%
\end{align}
so that the SET becomes%
\begin{equation}
T_{\mu\nu}=\frac{r_{0}}{2\kappa r^{3}}\left(  1-\mu\left(  r-r_{0}\right)
\right)  ^{\alpha-1}diag\left(  -2\alpha\mu r,-2,1-\mu\left(  r\left(
1-\alpha\right)  -r_{0}\right)  ,1-\mu\left(  r\left(  1-\alpha\right)
-r_{0}\right)  \right)  . \label{SETABG}%
\end{equation}
However, to have a vanishing $\rho\left(  r\right)  $ and $p_{t}(r)$ for
$r\geq\bar{r}$, we need $\alpha>1$. Note that for $\alpha=2$, we recover the
familiar shape function of the ABTW $\left(  \ref{b(r)AB}\right)  $. It is
easy to see that on the throat the SET $\left(  \ref{SETABG}\right)  $ becomes%
\begin{equation}
T_{\mu\nu}=\frac{1}{2\kappa r_{0}^{2}}diag\left(  -2\alpha\mu r_{0},-2,1+\mu
r_{0}\alpha,1+\mu r_{0}\alpha\right)
\end{equation}
and the inhomogenenous EoS function $\left(  \ref{o(r)ABG}\right)  $ reduces
to%
\begin{equation}
\omega\left(  r_{0}\right)  =\frac{1}{\alpha\mu r_{0}}, \label{omega}%
\end{equation}
while for $r=r_{0}+1/\mu$, one gets%
\begin{equation}
T_{\mu\nu}=diag\left(  0,0,0,0\right)  \qquad\alpha>1;\qquad\omega\left(
r_{0}+1/\mu\right)  =0.
\end{equation}
Regarding the energy density on the throat, one finds%
\begin{equation}
\rho\left(  r_{0}\right)  =-\frac{\alpha\mu}{\kappa r_{0}}%
\end{equation}
and if we impose the constraint $\left(  \ref{constr1}\right)  $, one finds%
\begin{equation}
\rho\left(  r_{0}\right)  =-\frac{\alpha\mu}{\kappa r_{0}}=-\frac{1}{\kappa
r_{0}^{2}}. \label{rhoGABr0}%
\end{equation}
By identifying $\left(  \ref{rhoGABr0}\right)  $ with the Casimir energy
density in $\left(  \ref{Tmn}\right)  $, one gets%
\begin{equation}
\rho\left(  r_{0}\right)  =-\frac{1}{\kappa r_{0}^{2}}=-\frac{\hbar c\pi^{2}%
}{720d^{4}}, \label{rhoI}%
\end{equation}
which implies%
\begin{equation}
r_{0}=\frac{3}{\pi}\sqrt{\frac{10}{\pi}}\frac{d^{2}}{l_{P}}\simeq
1.\,\allowbreak7\times10^{17}m \label{mu}%
\end{equation}
where we have fixed the plate separation at a distance of the order of
$10^{-9}m$. To have the exotic energy confined close to the throat, $\mu$ must
be huge, but the relationship $\left(  \ref{omega}\right)  $ constraints
$\alpha$ to be small. Therefore we conclude that even for the first GABTW
model, one finds the same problem of the assumption $\left(  \ref{a}\right)  $
and $\left(  \ref{aa}\right)  $, even if the size of the GABTW is estimated to
be $r_{0}\simeq1.\,\allowbreak7\times10^{17}m$, while for the original Casimir
size connected with $\left(  \ref{a}\right)  $ and $\left(  \ref{aa}\right)
$, the size of the wormhole throat was of the order $r_{0}\simeq10^{34}m$. For
this reason, we are going to consider this further generalization.

\subsection{Example II $\omega\left(  r\right)  =\left(  1-\mu\left(
r-r_{0}\right)  \right)  \left(  1-\nu\left(  r-r_{0}\right)  \right)
/r\left[  \alpha\mu\left(  \left(  1-\nu\left(  r-r_{0}\right)  \right)
-\beta\nu\left(  1-\mu\left(  r-r_{0}\right)  \right)  \right)  \right]  $}

\label{p2b}We now assume the following profile%
\begin{equation}
\omega\left(  r\right)  =\frac{\left(  1-\mu\left(  r-r_{0}\right)  \right)
\left(  1-\nu\left(  r-r_{0}\right)  \right)  }{r\left[  \alpha\mu\left(
\left(  1-\nu\left(  r-r_{0}\right)  \right)  -\beta\nu\left(  1-\mu\left(
r-r_{0}\right)  \right)  \right)  \right]  }, \label{o(r)GABTW}%
\end{equation}
where $\alpha,\beta\in%
\mathbb{R}
$ and $\mu,\nu$ are mass scales with $\mu>\nu$. In the range $r\in\left[
r_{0},\infty\right)  $, one finds%
\begin{equation}
\omega\left(  r\right)  \rightarrow\left\{
\begin{array}
[c]{cc}%
1/\left(  r_{0}\left(  \alpha\mu-\beta\nu\right)  \right)  & r\rightarrow
r_{0}\\
-1/\left(  \alpha+\beta\right)  & r\rightarrow\infty
\end{array}
\right.  .
\end{equation}
However, like in the Example I, it is of much more interest the following
assumption%
\begin{equation}
\omega\left(  r\right)  =0\qquad\mathrm{when}\qquad r=\bar{r}=r_{0}+\frac
{1}{\mu}.
\end{equation}
When this choice is adopted, from Eq.$\left(  \ref{form}\right)  $, one finds%
\begin{align}
b(r)  &  =r_{0}\,\exp\left[  -\int_{r_{0}}^{r}\,\frac{\left[  \alpha\mu\left(
\left(  1-\nu\left(  \bar{r}-r_{0}\right)  \right)  -\beta\nu\left(
1-\mu\left(  \bar{r}-r_{0}\right)  \right)  \right)  \right]  \bar{r}d\bar{r}%
}{\left(  1-\nu\left(  \bar{r}-r_{0}\right)  \right)  \left(  1-\mu\left(
\bar{r}-r_{0}\right)  \right)  \bar{r}}\right] \nonumber\\
&  =r_{0}\,\exp\left[  -\alpha\mu\int_{r_{0}}^{r}\,\frac{d\bar{r}}%
{1-\mu\left(  \bar{r}-r_{0}\right)  }+\beta\nu\int_{r_{0}}^{r}\,\frac{d\bar
{r}}{1-\nu\left(  \bar{r}-r_{0}\right)  }\right] \nonumber\\
&  =r_{0}\,\exp\left[  \alpha\ln\left(  1-\mu\left(  r-r_{0}\right)  \right)
-\beta\ln\left(  1-\nu\left(  r-r_{0}\right)  \right)  \right]  =r_{0}%
\frac{\left(  1-\mu\left(  r-r_{0}\right)  \right)  ^{\alpha}}{\left(
1-\nu\left(  r-r_{0}\right)  \right)  ^{\beta}}. \label{b(r)II}%
\end{align}
As we can see, the choice $\left(  \ref{o(r)ABG}\right)  $ leads to another
generalized ABTW, if we adopt also the following conditions%
\begin{align}
b(r)  &  =r_{0}\frac{\left(  1-\mu\left(  r-r_{0}\right)  \right)  ^{\alpha}%
}{\left(  1-\nu\left(  r-r_{0}\right)  \right)  ^{\beta}},\qquad
\Phi(r)=0;\qquad r_{0}\leq r\leq r_{0}+1/\mu\nonumber\\
b(r)  &  =0,\qquad\Phi(r)=0;\qquad r\geq r_{0}+1/\mu. \label{GABTWII}%
\end{align}
The components of the stress-energy tensor can be easily computed and
represented by Eq.$\left(  \ref{SET}\right)  $, while on the throat one finds%
\begin{equation}
T_{\mu\nu}=\frac{1}{2\kappa r_{0}^{2}}diag\left(  -2\left(  \alpha\mu-\beta
\nu\right)  r_{0},-2,1+r_{0}\left(  \alpha\mu-\beta\nu\right)  ,1+r_{0}\left(
\alpha\mu-\beta\nu\right)  \right)  \label{SETABGIIr0}%
\end{equation}
and for $r\geq r_{0}+1/\mu$, the whole SET is vanishing, namely a Minkowski
SET. Note that for $\alpha=2$ and $\beta=0$, we recover the familiar shape
function of the ABTW $\left(  \ref{b(r)AB}\right)  $. On the throat, we can
impose that the energy density be%
\begin{equation}
\rho\left(  r_{0}\right)  =-\frac{1}{\kappa r_{0}}\left(  \alpha\mu-\beta
\nu\right)  =-\frac{\hbar c\pi^{2}}{720d^{4}} \label{rhoid}%
\end{equation}
and we the additional condition $\left(  \ref{constr1}\right)  $, one finds%
\begin{equation}
\rho\left(  r_{0}\right)  =\frac{1}{\kappa r_{0}^{2}}=\frac{\hbar c\pi^{2}%
}{720d^{4}}%
\end{equation}
leading to%
\begin{equation}
r_{0}=\frac{3d^{2}}{\pi l_{P}}\sqrt{\frac{10}{\pi}}, \label{r0II}%
\end{equation}
which is in agreement with the result $\left(  \ref{mu}\right)  $. Note that
in proximity of the throat, the shape function $\left(  \ref{b(r)II}\right)  $
can be rewritten as%
\begin{equation}
b\left(  r\right)  \simeq r_{0}\left(  1-\left(  \alpha\mu-\beta\nu\right)
\left(  r-r_{0}\right)  \right)
\end{equation}
which looks like the shape function $\left(  \ref{b(r)Cr0}\right)  $. Note
also that the constraint $\left(  \ref{constr1}\right)  $ leads to%
\begin{equation}
\mu=\frac{\beta}{\alpha}\nu+\frac{\pi l_{P}}{3\alpha d^{2}}\sqrt{\frac{\pi
}{10}}%
\end{equation}
and by setting%
\begin{equation}
\alpha=\beta+1;\qquad\beta=\beta\text{ \qquad and\qquad}\nu\gg\frac{1}{r_{0}},
\label{ab1}%
\end{equation}
we can mimic the shape function $\left(  \ref{b(r)Cr0}\right)  $. However,
this time the parameter $\mu$ can be large satisfying therefore the request of
concentrating the exotic matter close to the throat. One can see that although
the original Casimir structure for the SET $\left(  \ref{TTmn}\right)  $ is
reproduced, the whole SET is divided by a factor 2 while the SET $\left(
\ref{SET}\right)  $ becomes on the throat
\begin{equation}
\rho\left(  r_{0}\right)  diag\left(  1,1,-1,-1\right)  ,
\end{equation}
which is in agreement with the Casimir SET except for the radial pressure.
This behavior was present also in the SET of the Ref.\cite{EPJC}.

\section{Reexamining the original Casimir structure}

\label{p3}In this section we are going to reconsider the shape function
$\left(  \ref{b(r)C}\right)  $ without the approximation leading to the form
$\left(  \ref{b(r)Cr0}\right)  $. We have already observed that, for $r\gg
r_{0}$, the shape function assumes a de Sitter profile. However, we can also
observe that $b\left(  r\right)  $ without approximations has one real root.
This is located at%
\begin{equation}
b\left(  \bar{r}\right)  =0\qquad\Longleftrightarrow\qquad\bar{r}%
=r_{0}\sqrt[3]{1+\frac{3}{r_{0}^{2}\rho_{0}\kappa}}.
\end{equation}
Therefore, it is straightforward to assume that%
\begin{align}
b(r)  &  =r_{0}-\frac{\rho_{0}\kappa}{3}\left(  r^{3}-r_{0}^{3}\right)
,\qquad\Phi(r)=0;\qquad r_{0}\leq r\leq\bar{r}\nonumber\\
b(r)  &  =0,\qquad\Phi(r)=0;\qquad r\geq\bar{r}. \label{CbP}%
\end{align}
Nevertheless, this choice does not lead to a Minkowski space outside the
region located at $r=\bar{r}$, because the SET has a structure which looks
like the SET\ of Eq.$\left(  \ref{TmnO}\right)  $. However, it is immediate to
realize that the profile described by $\left(  \ref{CbP}\right)  $ can be
generalized to%
\begin{align}
b(r)  &  =\frac{1}{r_{0}^{\alpha-1}}\left[  r_{0}-\frac{\rho_{0}\kappa
}{3\alpha}\left(  r^{3}-r_{0}^{3}\right)  \right]  ^{\alpha},\qquad
\alpha>1;\qquad\Phi(r)=0;\qquad r_{0}\leq r\leq\bar{r}\nonumber\\
b(r)  &  =0,\qquad\Phi(r)=0;\qquad r\geq\bar{r}, \label{GCbP}%
\end{align}
where, this time%
\begin{equation}
\bar{r}=r_{0}\sqrt[3]{1+\frac{3\alpha}{r_{0}^{2}\rho_{0}\kappa}}. \label{Sol}%
\end{equation}
In this way by imposing the EoS we find%
\begin{align}
\omega\left(  r\right)   &  =\frac{1}{r^{3}\rho_{0}\kappa}\left(  r_{0}%
-\frac{\rho_{0}\kappa}{3\alpha}\left(  r^{3}-r_{0}^{3}\right)  \right)
\nonumber\\
\omega\left(  r\right)   &  =0\qquad r\geq\bar{r},\qquad\qquad\omega\left(
r_{0}\right)  =\frac{1}{r_{0}^{2}\rho_{0}\kappa}.
\end{align}
The corresponding SET can be derived from the Eq.$\left(  \ref{TmnDec}\right)
$ where%
\begin{equation}
\rho\left(  r\right)  =-\frac{\rho_{0}}{r_{0}^{\alpha-1}}\left[  r_{0}%
-\frac{\rho_{0}\kappa}{3\alpha}\left(  r^{3}-r_{0}^{3}\right)  \right]
^{\alpha-1}\qquad\Longrightarrow\qquad\rho\left(  r_{0}\right)  =-\rho_{0}.
\end{equation}
This is very interesting, because independently on the exponent $\alpha$, on
the throat, we can find always the Casimir energy density. A result different
compared with $\left(  \ref{rhoI}\right)  $ and $\left(  \ref{rhoid}\right)
$. Note that outside the region located at $r=\bar{r}$, the spacetime is
Minkowski. Moreover by fixing%
\begin{equation}
\omega\left(  r_{0}\right)  =\frac{1}{r_{0}^{2}\rho_{0}\kappa}=1,
\label{o(r)C}%
\end{equation}
we can recover the Casimir structure of the SET and putting numbers inside the
previous relationship, one finds%
\begin{equation}
r_{0}=\sqrt{\frac{1}{\rho_{0}\kappa}}=\frac{3d^{2}}{\pi l_{P}}\sqrt{\frac
{10}{\pi}},
\end{equation}
which is compatible with what we have investigated in section \ref{p2}. To
determine if the root $\left(  \ref{Sol}\right)  $ is close to the throat, we
need to evaluate the ratio%
\begin{align}
R  &  =\frac{\bar{r}-r_{0}}{r_{0}}=\sqrt[3]{1+\frac{3\alpha}{r_{0}^{2}\rho
_{0}\kappa}}-1=\sqrt[3]{1+3\alpha}-1\\
&  \Longrightarrow%
\begin{array}
[c]{cc}%
0.59 & \alpha=1\\
0.91 & \alpha=2
\end{array}
,
\end{align}
where we have used the constraint $\left(  \ref{o(r)C}\right)  $. To have
consistency, $1<\alpha\leq2$. For $\alpha>2$, $R>1$: in this case the exotic
matter is not confined close tho the throat. Note that it is sufficient to
choose $\alpha$ very close to $1$ to have the Minkowski space time outside
$\bar{r}$. Unfortunately as we can see $R<1$ and not $R\ll1$. This is due to
the constraint induced by $\left(  \ref{o(r)C}\right)  $. If we give up this
constraint, we cannot recover the Casimir SET structure.

\section{Conclusions}

\label{p4}In this paper, we have further extended the study began by Morris
and Thorne\cite{MT}; Morris, Thorne and Yurtsever in Ref.\cite{MTY} and,
subsequently explored by Visser\cite{Visser} on the Casimir effect as a
possible source for a TW. We have also further extended the results obtained
in Ref.\cite{EPJC} and the result of such an extension has revealed an
interesting further connection between the Casimir source and a particular TW:
an ABTW. However, the structure of the solution directly connected with the
Casimir source presents some problems in reproducing the features of an ABTW.
One of these problems is the lack of a smooth change between the curved and
flat space. For this reason, we have investigated some profiles, termed GABTW,
that in some particular cases can mimic the original solution derived by the
Casimir source. The first GABTW profile failed to be a good candidate because
the constraint $\left(  \ref{constr1}\right)  $ and the relationship $\left(
\ref{omega}\right)  $ forbid to fix large values of the parameter $\mu$, a
necessary request to have the exotic matter confined close to the throat. For
this reason, we have examined another profile defined by Eq.$\left(
\ref{b(r)II}\right)  $, obtained by imposing the inhomogeneous EoS $\left(
\ref{o(r)GABTW}\right)  $. This profile has produced interesting results which
go in a completely opposite direction with respect to the results obtained in
Ref.\cite{EPJC}: namely in this paper the wormhole throat can be huge, while
in Ref.\cite{EPJC} is Planckian. The main difference in this result is that,
in case of Ref.\cite{EPJC}, the Casimir source has a plate separation which is
variable, while in this paper it is not: it is a parameter. A question is in
order: why do we insist in analyzing an ABTW and its generalization? Because,
by definition, the region of the exotic matter is confined in a very small
region and outside of this region, space-time is flat. Therefore, instead of
having a long tail that asymptotically becomes flat, the flatness is
\textit{almost} in proximity of the throat. Even if the results related to the
profiles $\left(  \ref{b(r)ABG}\right)  $ and $\left(  \ref{b(r)II}\right)  $
seem not to be encouraging, we have to observe that the estimated size of the
wormhole throat obtained in Eq.$\left(  \ref{aa}\right)  $ predicts a wormhole
throat of the order of $10^{34}m$, but the GABTW described by $\left(
\ref{b(r)ABG}\right)  $ and $\left(  \ref{b(r)II}\right)  $ predicts a
wormhole throat of the order of $10^{17}m$. This huge size is essentially due
to the request of having imposed $\omega\left(  r_{0}\right)  =1$. Note also
that the size $10^{17}m$ is principally due to the plate separation of the
order of the $nm$, which is the actual plate separation used in the
experiments, but an interesting improvement can arrive at the next scale,
namely a $pm$ scale, which is surely more easier to reach compared to the $fm$
scale. In this case, one obtains%
\begin{equation}
r_{0}\simeq10^{11}m.
\end{equation}
Note also that the presence of the Planck length square in the expression of
the wormhole throat is principally due to the combination of the Newton's
constant $G$, the Planck constant $\hbar$ and the speed of light $c$. These
last two constants appear in the Casimir energy density calculation. It is
interesting to note that in some Casimir experiments, if the plates enter in a
superconductive phase, it is possible to show an increase of negative energy
\cite{EC}. This is promising because from Eq.$\left(  \ref{r0II}\right)  $,
one finds%
\begin{equation}
r_{0}=\frac{3d^{2}}{\pi l_{P}}\sqrt{\frac{10}{A\pi}},
\end{equation}
where $A$ comes from the increase of the negative energy density coming from
the superconducting phase. This means that in this particular situation it
could be possible to combine the plates separation with the superconducting
phase energy density increase to obtain a more realistic wormhole throat size.
Finally, we have investigated also the profile generated by the Casimir energy
source without approximation, but having in mind a GABTW structure. There are
several interesting points about the profile $\left(  \ref{GCbP}\right)  $:
the first one is that the relationships $\left(  \ref{mu}\right)  $ and
$\left(  \ref{r0II}\right)  $ are confirmed even for this shape function, the
second one is the reproduction of the Casimir source for every exponent
$\alpha$. Therefore, it seems that with the profile $\left(  \ref{GCbP}%
\right)  $, everything seems to go in the right ballpark except for the exotic
matter region that cannot be shrunk to a very small region exactly like for
the profile $\left(  \ref{b(r)II}\right)  $. At this stage, we do not know how
much is important to force the exotic matter to stay in a region very close to
the throat to keep the GABTW $\left(  \ref{GCbP}\right)  $ valid and, at the
same time, the reproduction of some features of the Casimir SET, i.e.
traceless and divergenceless. To conclude, we have also to point out that in
the context of \textit{Self-Sustained Traversable Wormholes}, namely TW
sustained by their own quantum
fluctuations\cite{RG,RG1,RGFSNL,RGFSNL1,RGFSNL2}, could be interesting to
consider how the Casimir source and the quantum fluctuation carried by the
graviton combine to see if the GABTW can be self-sustained in this context. In
this picture, the Casimir source could be interpreted as the switch on to
power the traversability of the wormhole.

\appendix{}

\section{Features of the Traversable Wormhole of subesction\ref{p2b}}

\label{p5}In this section we are going to explore some of the features of the
GABTW $\left(  \ref{b(r)II}\right)  $. The motivation of examining only this
kind of profile is that it is quite general to include many GABTW profiles in
proximity of the throat. We begin to examine the proper length which is
defined as%
\begin{equation}
l\left(  r\right)  =\pm\int_{r_{0}}^{r}\frac{dr^{\prime}}{\sqrt{1-\frac
{b\left(  r^{\prime}\right)  }{r^{\prime}}}},
\end{equation}
where the\textquotedblleft$\pm$\textquotedblright\ depends on the wormhole
side we are. In this case, one gets%
\begin{equation}
l\left(  r\right)  =\pm\int_{r_{0}}^{r}\frac{dr^{\prime}}{\sqrt{1-\frac{r_{0}%
}{r^{\prime}}\frac{\left(  1-\mu\left(  r^{\prime}-r_{0}\right)  \right)
^{\alpha}}{\left(  1-\nu\left(  r^{\prime}-r_{0}\right)  \right)  ^{\beta}}}}.
\label{l(r)}%
\end{equation}
The exact evaluation of the integral is really complicated. However, it is
sufficient to consider that the amount of exotic matter for the GABTW is
concentrated near the throat by construction. Therefore, it is sufficient to
consider the expression close to the throat with $\mu$ very large. To this
purpose, we can write%
\begin{align}
&  l\left(  r\right)  \underset{r\rightarrow r_{0}}{\simeq}\pm\frac{1}%
{\sqrt{1+\alpha\mu r_{0}-\beta\nu r_{0}}}\int_{r_{0}}^{r}\frac{\sqrt
{r^{\prime}}dr^{\prime}}{\sqrt{r^{\prime}-r_{0}}}\nonumber\\
&  =\pm\frac{1}{\sqrt{1+\alpha\mu r_{0}-\beta\nu r_{0}}}\left[  \sqrt{r}%
\sqrt{r-r_{0}}+r_{0}{\ln\left(  \sqrt{\frac{r}{r_{0}}}+\sqrt{\frac{r}{r_{0}%
}-1}\right)  }\right]  \qquad\qquad r_{0}\leq r\leq r_{0}+1/\mu
\end{align}
and the complete $l\left(  r\right)  $ is%
\begin{equation}
l\left(  r\right)  =\pm\left\{
\begin{array}
[c]{cc}%
\frac{1}{\sqrt{1+\alpha\mu r_{0}-\beta\nu r_{0}}}\left[  \sqrt{r}\sqrt
{r-r_{0}}+r_{0}{\ln\left(  \sqrt{\frac{r}{r_{0}}}+\sqrt{\frac{r}{r_{0}}%
-1}\right)  }\right]  & r_{0}\leq r\leq r_{0}+1/\mu\\
\frac{1}{\sqrt{1+\alpha\mu r_{0}-\beta\nu r_{0}}}\left[  \sqrt{r_{0}+\frac
{1}{\mu}}\sqrt{\frac{1}{\mu}}+r_{0}{\ln\left(  \sqrt{1+\frac{1}{r_{0}\mu}%
}+\sqrt{\frac{1}{r_{0}\mu}}\right)  }\right]  +r-\left(  r_{0}+\frac{1}{\mu
}\right)  & r\geq r_{0}+1/\mu
\end{array}
\right\}  .
\end{equation}
Note that the time lapse $dt$, and proper time lapse as measured by the
observer $d\tau$, for the GABTW are the same, because the redshift function is
nought. In a similar way, to compute the embedded surface, we need to evaluate%

\begin{equation}
z\left(  r\right)  =\pm\int_{r_{0}}^{r}\frac{dr^{\prime}}{\sqrt{\frac
{r^{\prime}}{b\left(  r^{\prime}\right)  }-1}},
\end{equation}
which, for the present case, is%
\begin{equation}
z\left(  r\right)  =\pm\int_{r_{0}}^{r}\frac{dr^{\prime}}{\sqrt{\frac
{r^{\prime}}{r_{0}}\frac{\left(  1-\nu\left(  r^{\prime}-r_{0}\right)
\right)  ^{\beta}}{\left(  1-\mu\left(  r^{\prime}-r_{0}\right)  \right)
^{\alpha}}-1}}.
\end{equation}
By repeating the same procedure adopted for the proper length, we can write%
\begin{equation}
z\left(  r\right)  \underset{r\rightarrow r_{0}}{\simeq}\pm\frac{\sqrt{r_{0}}%
}{\sqrt{1+\alpha\mu r_{0}-\beta\nu r_{0}}}\int_{r_{0}}^{r}\frac{dr^{\prime}%
}{\sqrt{r^{\prime}-r_{0}}}=\pm\frac{2\sqrt{r_{0}}\sqrt{r-r_{0}}}%
{\sqrt{1+\alpha\mu r_{0}-\beta\nu r_{0}}}.
\end{equation}
To further investigate the properties of the GABTW, we consider the
computation of the total gravitational energy for a wormhole\cite{NZCP},
defined as%
\begin{equation}
E_{G}\left(  r\right)  =\int_{r_{0}}^{r}\left[  1-\sqrt{\frac{1}{1-b\left(
r^{\prime}\right)  /r^{\prime}}}\right]  \rho\left(  r^{\prime}\right)
dr^{\prime}r^{\prime2}+\frac{r_{0}}{2G}=M-M_{\pm}^{P},\label{Eg}%
\end{equation}
where $M$ is the total mass and $M^{P}$ is the proper mass, respectively.
Differently from the case where the Casimir energy was considered variable
depending on the radial radius $r$, here we have no asymptotic mass, since
outside the radius $\mu^{-1}$, spacetime is flat. In particular we find for
the total mass%
\begin{align}
Mc^{2} &  =\int_{r_{0}}^{r_{0}+\frac{1}{\mu}}4\pi\rho\left(  r^{\prime
}\right)  r^{\prime2}dr^{\prime}=\frac{4\pi}{3}\left[  \left(  r_{0}+\frac
{1}{\mu}\right)  ^{3}-r_{0}^{3}\right]  \left(  \frac{\hbar c\pi^{2}}%
{720d^{4}}\right)  \label{M}\\
&  \underset{r\rightarrow r_{0}}{\simeq}\frac{4\pi r_{0}^{2}}{\mu}\left(
\frac{\hbar c\pi^{2}}{720d^{4}}\right)  ,
\end{align}
where we have considered the Casimir energy density as a source and we have
momentarily reintroduced the speed of light. For the proper mass, one gets%
\begin{align}
M_{\pm}^{P}c^{2} &  =\pm\int_{r_{0}}^{r_{0}+\frac{1}{\mu}}\frac{4\pi
\rho\left(  r^{\prime}\right)  r^{\prime2}}{\sqrt{1-b\left(  r^{\prime
}\right)  /r^{\prime}}}dr^{\prime}\underset{r\rightarrow r_{0}}{\simeq}%
\pm\frac{4\pi r_{0}^{2}}{\sqrt{1+\alpha\mu r_{0}-\beta\nu r_{0}}}\left(
\frac{\hbar c\pi^{2}}{720d^{4}}\right)  \int_{r_{0}}^{r_{0}+\frac{1}{\mu}%
}\frac{dr^{\prime}}{\sqrt{r-r_{0}}}\nonumber\\
&  \simeq\pm\frac{8\pi r_{0}^{2}\sqrt{r_{0}}}{\sqrt{\mu}\sqrt{1+\alpha\mu
r_{0}-\beta\nu r_{0}}}\left(  \frac{\hbar c\pi^{2}}{720d^{4}}\right)
\qquad\underset{\mu r_{0}\gg1}{\longrightarrow}\qquad\pm\frac{8\pi r_{0}^{2}%
}{\mu\sqrt{\alpha}}\left(  \frac{\hbar c\pi^{2}}{720d^{4}}\right)  ,\label{Mp}%
\end{align}
where the \textquotedblleft$\pm$\textquotedblright\ depends one the wormhole
side we are. Thus the total gravitational energy $\left(  \ref{Eg}\right)  $
becomes%
\begin{equation}
E_{G}\left(  r\right)  \simeq\left(  \frac{\hbar c\pi^{2}}{720d^{4}}\right)
\frac{4\pi r_{0}^{2}}{\mu}\left[  1\pm\frac{2}{\sqrt{\alpha}}\right]
\end{equation}
For large $\mu$, one finds that $E_{G}$ vanishes, independently on the scale
choice we make about $\mu$. Another important traversability condition is that
the acceleration felt by the traveller should not exceed Earth's gravity
$g_{\oplus}\simeq980$ $cm/s^{2}$. In an orthonormal basis of the traveller's
proper reference frame, we can find%
\begin{equation}
\left\vert \mathbf{a}\right\vert =\left\vert \sqrt{1-\frac{b\left(  r\right)
}{r}}e^{-\Phi(r)}\left(  \gamma e^{\Phi(r)}\right)  ^{\prime}\right\vert
\leq\frac{g_{\oplus}}{c^{2}}%
\end{equation}
and in this case, because $\Phi(r)=0$, the traveller has no acceleration,
which is in agreement with Ref.\cite{MT}. As regards the lateral tidal forces,
we find%
\begin{gather}
\left\vert \frac{\gamma^{2}c^{2}}{2r^{2}}\left[  \frac{v^{2}\left(  r\right)
}{c^{2}}\left(  b^{\prime}\left(  r\right)  -\frac{b\left(  r\right)  }%
{r}\right)  +2r\left(  r-b\left(  r\right)  \right)  \Phi^{\prime}\left(
r\right)  \right]  \right\vert \left\vert \eta\right\vert \nonumber\\
=\left\vert \frac{\gamma^{2}c^{2}}{2r^{3}}\left[  -\frac{v^{2}\left(
r\right)  b\left(  r\right)  }{c^{2}}\left(  \frac{1}{\omega\left(  r\right)
}+1\right)  \right]  \right\vert \left\vert \eta\right\vert \leq g_{\oplus
},\label{LTC}%
\end{gather}
where we have used the relationship $\left(  \ref{o(r)}\right)  $. This is a
constraint about the velocity with which observers traverse the wormhole.
$\eta$ represents the size of the traveller which can be fixed approximately
equal, at the symbolic value of $2$ $m$\cite{MT}. If we assume a constant
speed $v$ and $\gamma\simeq1$, close to the throat, the lateral tidal
constraint becomes%
\begin{gather}
\left\vert \frac{\gamma^{2}c^{2}}{2r_{0}^{2}}\left[  \frac{v^{2}\left(
r_{0}\right)  }{c^{2}}\left(  -r_{0}\left(  \alpha\mu-\beta\nu\right)
-1\right)  \right]  \right\vert \left\vert 2\right\vert \simeq\left\vert
\left[  \frac{v^{2}\left(  r_{0}\right)  }{r_{0}^{2}}\right]  \right\vert
\lesssim g_{\oplus}\nonumber\\
\qquad\Longrightarrow\qquad v\lesssim r_{0}\sqrt{g_{\oplus}}\qquad
\Longrightarrow\qquad v\lesssim3.1r_{0}\text{ }m/s.\label{LTCt}%
\end{gather}
If the observer has a vanishing $v$, then the tidal forces are null. We can
use these last estimates to complete the evaluation of the crossing time which
approximately is%
\begin{equation}
\Delta t\simeq2\times10^{4}\frac{3r_{0}}{4v}\simeq5\times10^{3}s,
\end{equation}
which is in agreement with the estimates found in Ref.\cite{MT}.The last
property we are going to discuss is the \textquotedblleft total
amount\textquotedblright\ of ANEC violating matter in the spacetime\cite{VKD}
which is described by Eq. $\left(  \ref{IV}\right)  $. For the metric $\left(
\ref{b(r)II}\right)  $, one obtains%
\begin{align}
I_{V} &  =\frac{1}{\kappa}\int_{r_{0}}^{r_{0}+\frac{1}{\mu}}\left(  r-b\left(
r\right)  \right)  \left[  \ln\left(  \frac{e^{2\phi(r)}}{1-\frac{b\left(
r\right)  }{r}}\right)  \right]  ^{\prime}dr=\frac{1}{\kappa}\int_{r_{0}%
}^{r_{0}+\frac{1}{\mu}}\left(  \frac{b\left(  r\right)  }{r}-b^{\prime}\left(
r\right)  \right)  dr\label{IV}\\
&  \simeq\frac{1}{\kappa}\left[  \int_{r_{0}}^{r_{0}+\frac{1}{\mu}}\alpha\mu
r_{0}-\beta\nu r_{0}+1+c\left(  r\right)  \left(  r-r_{0}\right)  \right]  dr,
\end{align}
where we have approximated the expression close to the throat and where we
have defined%
\begin{equation}
c\left(  r\right)  =-\alpha\mu-r_{0}^{-1}+\beta\nu-{\alpha}^{2}{\mu}^{2}%
r_{0}+\alpha\mu r_{0}\beta\,\nu+\alpha{\mu}^{2}r_{0}-\left(  -\alpha\mu
r_{0}+\beta\nu r_{0}\right)  \beta\nu-\beta{\nu}^{2}r_{0}.
\end{equation}
After the integration, we find%
\begin{align}
I_{V} &  =\frac{1}{\kappa}\left(  \frac{3}{2}{\alpha}r_{0}-\frac{1}{2}{\alpha
}^{2}r_{0}+{\frac{\nu r_{0}\beta\alpha}{\mu}}-{\frac{\beta\nu r_{0}}{\mu}%
}-{\frac{\alpha}{2\mu}}+\frac{1}{\mu}\right.  \nonumber\\
&  \left.  -{\frac{\beta{\nu}^{2}r_{0}}{2{\mu}^{2}}}-{\frac{{\beta}^{2}{\nu
}^{2}r_{0}}{2{\mu}^{2}}}+{\frac{\beta\nu}{2{\mu}^{2}}}-{\frac{1}{2{\mu}%
^{2}r_{0}}}\right)  \underset{\mu r_{0}\gg1}{\simeq}\frac{1}{\kappa}\left(
\frac{3}{2}{\alpha}r_{0}-\frac{1}{2}{\alpha}^{2}r_{0}\right)  ,
\end{align}
and the result is finite. Therefore we can conclude that, in proximity of the
throat the ANEC can be arbitrarily small as it should be.

\end{document}